\documentclass[aps,twocolumn,
superscriptaddress,
footinbib,
pra]{revtex4-1}

\usepackage{amsmath,amssymb,bm,bbold}
\usepackage{mathtools}
\usepackage{graphicx}
\usepackage{multirow}
\usepackage{epstopdf}
\usepackage{latexsym}
\usepackage[normalem]{ulem} 
\usepackage[caption=false]{subfig}
\usepackage[usenames,dvipsnames]{color}
\usepackage{hyperref}
\usepackage{xcolor}
\usepackage{braket}
\usepackage{physics}
\usepackage{stackengine}
\usepackage[utf8]{inputenc} 
\usepackage{natbib}

\begin{document}

\title{Semiclassical simulations predict glassy dynamics for disordered Heisenberg models}
\date{\today}

\author{P.~Schultzen}
\email{These authors contributed equally to this work. }
\author{T.~Franz}
\email{These authors contributed equally to this work. }
\author{C.~Hainaut}
\email{These authors contributed equally to this work. }
\author{S.~Geier}
\author{A.~Salzinger}
\author{A.~Tebben}
\author{G.~Z\"urn}
\affiliation{Physikalisches Institut, Universit\"at Heidelberg, Im Neuenheimer Feld 226, 69120 Heidelberg, Germany}
\author{M.~G\"{a}rttner}
\affiliation{Physikalisches Institut, Universit\"at Heidelberg, Im Neuenheimer Feld 226, 69120 Heidelberg, Germany}
\affiliation{Kirchhoff-Institut f\"{u}r Physik, Universit\"{a}t Heidelberg, Im Neuenheimer Feld 227, 69120 Heidelberg, Germany}
\affiliation{Institut f\"ur Theoretische Physik, Ruprecht-Karls-Universit\"at Heidelberg, Philosophenweg 16, 69120 Heidelberg, Germany}
\author{M.~Weidemüller}
\email{Corresponding author. weidemueller@uni-heidelberg.de}
\affiliation{Physikalisches Institut, Universit\"at Heidelberg, Im Neuenheimer Feld 226, 69120 Heidelberg, Germany}
\begin{abstract}
We numerically study out-of-equilibrium dynamics in a family of Heisenberg models with $1/r^6$ power-law interactions and positional disorder. Using the semi-classical discrete truncated Wigner approximation (dTWA) method, we investigate the time evolution of the magnetization and ensemble-averaged single-spin purity for a strongly disordered system after initializing the system in an out-of-equilibrium state. We find that both quantities display robust glassy behavior for almost any value of the anisotropy parameter of the Heisenberg Hamiltonian. Furthermore, a systematic analysis allows us to quantitatively show that, for all the scenarios considered, the stretch power lies close to the one analytically obtained in the Ising limit. This indicates that glassy relaxation behavior occurs widely in disordered quantum spin systems, independent of the particular symmetries and integrability of the Hamiltonian.
\end{abstract}

\maketitle  
\section{Introduction}
Recent experimental progress and the development of more advanced numerical tools allow the exploration of far-from equilibrium physics of many-body systems, e.g. transport \cite{Jepsen2020,PhysRevLett.120.070501}, localization \cite{annurev-conmatphys-031214-01472,doi.org/10.1002/andp.201770053} and dynamical phase transitions \cite{PhysRevLett.119.080501}. In disordered quantum systems, peculiarly rich relaxation dynamics has been found where the interplay of interactions and randomness results in new and intrinsically non-equilibrium effects such as pre-thermalization \cite{Eigen2018,Gring1318}, quantum scars \cite{Turner2018} and aging \cite{PhysRevLett.88.257202}.

Strikingly, a large variety of disordered materials \cite{Binder1986a, Phillips_1996,Gotze_1992,doi:10.1063/1.466117} feature slow sub-exponential relaxation behaviors. Phenomenologically, these can be described by a stretched exponential law $\exp(-(t / \tau)^{\beta})$ with relaxation time $\tau$ and stretch power $\beta$~\cite{1854AnP...167...56K}. Despite the huge success of this law in capturing relaxation dynamics of disordered systems, derivations of this function from microscopic
models are rare \cite{Phillips_1996}. 
By exploring relaxation behaviours of three different classical disordered models, Klafter and Shlesinger \cite{Klafter1986OnSystems} concluded that the scale-invariant property of the relaxation timescales is the underlying feature resulting in the appearance of stretched exponential law. Similarly, this behavior has also been observed in a disordered open quantum system where the decay rates, coupling the system to an environment, display scale-invariance \cite{Choi2017e,Kucsko2018b}.

In contrast to dissipative systems, closed quantum systems are subjected to unitary dynamics where relaxation can solely be explained by interactions. In recent work, we have derived analytically the occurrence of stretched exponential law for the disordered quantum Ising model in the thermodynamic limit, where the interactions between different spins feature a scale-invariant distribution \cite{schultzen2021glassy}. These observations indicate also in disordered \emph{quantum} systems that scale-invariance implies glassy dynamics. Scrutinizing this hypothesis requires studying more general classes of disordered quantum spin systems.
A suitable candidate for this is the XXZ Heisenberg model for which we recently observed glassy dynamics in an experiment with Rydberg atoms \cite{Signoles2021}. Nevertheless, in contrast to the integrable quantum Ising model the XXZ Heisenberg model is in general non-integrable, which prevents direct analytical derivations beyond the disorder-free one-dimensional case \cite{Franchini_2017}. Furthermore, due to the exponentially increasing size of the Hilbert space, exact numerical studies are limited to small system sizes. Among different numerical methods, the semiclassical method of dTWA \cite{PhysRevX.5.011022} already succeeded to capture the glassy behavior observed in a Rydberg spin experiment \cite{Signoles2021}. In addition this method becomes exact for describing the evolution of the magnetization in the Ising limit \cite{PhysRevX.5.011022}. These arguments imply that this method is a natural candidate to address the question of glassy dynamics of  disordered systems composed of a large number of particles.

Within this work, we use the dTWA method (see Appendix \ref{dtwa} for details on the numerical method and benchmark calculations) to investigate the occurrence of glassy dynamics for the relaxation of magnetization and ensemble-averaged single-spin purity in the general XYZ Heisenberg model. The article is organized as follows: In section \ref{section1} we introduce the system and the relevant observable, in section \ref{section2} we focus on the dynamics of the particular scenario of the XXZ Heisenberg model. In section \ref{section3} this investigation is extended to the XYZ Hamiltonian. 


\section{XYZ Heisenberg model}\label{section1}

We investigate the dynamics of $N$ spin-1/2 particles described by the general XYZ Hamiltonian 
\begin{equation}
\label{Eq:XYZ}
    H = \frac{1}{2} \sum_{i, j} \frac{J_{ij}}{|J|}\left( J_x \hat{\sigma}_x^{i}\hat{\sigma}_x^{j} + J_y \hat{\sigma}_y^{i} \hat{\sigma}_y^{j} + \Delta \hat{\sigma}_z^{i} \hat{\sigma}_z^{j}  \right) \, ,
\end{equation}
where for each spin $i$ the Pauli matrices $\hat{\sigma}^i_\alpha$ with $\alpha \in \{ x,y,z\}$ are introduced. The spins interact via Van der Waals interactions $J_{ij} = 1/r_{ij}^6$ with $r_{ij}$ being the distance between spins $i$ and $j$. The prefactors $J_x$, $J_y$ and $\Delta$ describe the contributions of the different interaction terms and are normalized by $|J| = (J_x^2 + J_y^2 + \Delta ^2)^{1/2}$. In the case of $J_x = J_y = \Delta = 1$, the system is SU(2) symmetric which implies a conservation of all the three components of the global magnetization: $\overline{\Braket{\hat{\sigma}_x}}$, $\overline{\Braket{\hat{\sigma}_y}}$ and $\overline{\Braket{\hat{\sigma}_z}}$  where $  \overline{\Braket{\hat{\sigma}_\alpha}}$ are defined as
\begin{equation}
   \overline{\Braket{\hat{\sigma}_\alpha}} = \frac{1}{N }\sum_{i=1}^N \Braket{\hat{\sigma}^{i}_\alpha} \, .
\end{equation}
As this peculiar isotropic case displays no relaxation dynamics, we will exclude it from numerical investigations in the following. 
In the case of $J_x = J_y = 1$ the model of Eq.~\ref{Eq:XYZ} reduces to the anisotropic XXZ Heisenberg model featuring U(1) symmetry.
Finally, we note that in the limits $\Delta \rightarrow \pm \infty$ the (anti)ferromagnetic quantum Ising model is recovered. 

In this paper, we numerically study the dynamics of an ensemble of spins homogeneously distributed in three dimensions. We focus on the dynamics of the transversal magnetization $\overline{\Braket{\hat{\sigma}_x}}$ and the ensemble-averaged single-spin purity 
\begin{equation}
    \overline{\mathrm{tr}(\rho)^2} = \frac{1}{2}\left(1 + \overline{\Braket{ \hat{\sigma}_x }^2} + \overline{\Braket{ \hat{\sigma}_y }^2} + \overline{\Braket{ \hat{\sigma}_z }^2}\right) \,
\end{equation}
after having initialized the system in the state $\ket{\Psi_0}=\ket{\rightarrow}^N$ where $\ket{\rightarrow}$ is the $\hat{\sigma}_x$ eigenstate with positive eigenvalue. 
The purity quantifies the entanglement between each single spin and its environment. The purity of a single spin can 
take values of $1/2 \leq  \overline{\mathrm{tr}(\rho)^2} \leq 1 $ and is directly connected to the second order Rényi entropy $S_2=-\log( \mathrm{tr}[\rho^2])$.

\section{Stretched exponential relaxation in the Heisenberg XXZ model}\label{section2}

\begin{figure}[h]
\includegraphics[width= \linewidth]{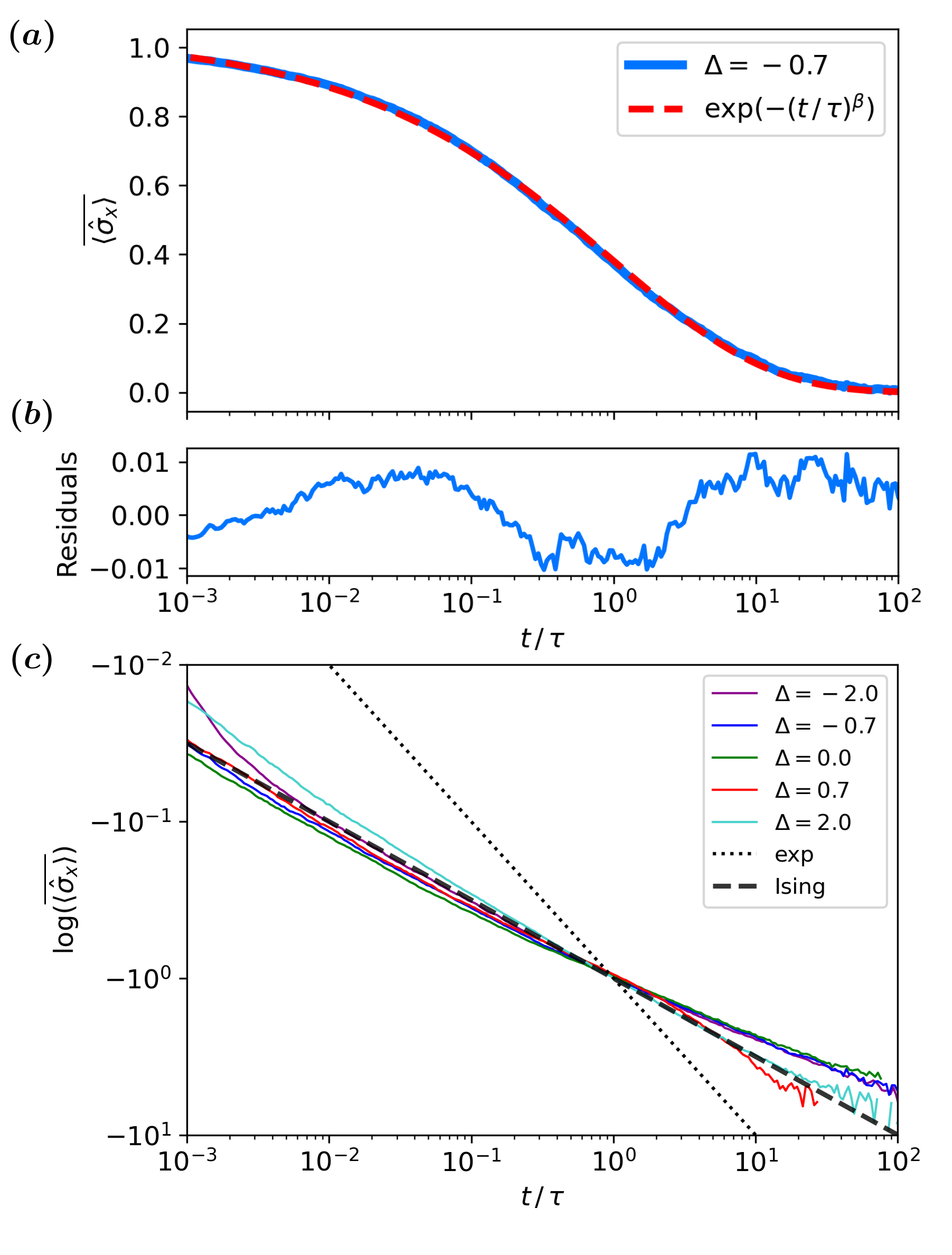}
\caption{ \textbf{(a)} Exemplary relaxation dynamics of the transverse magnetization $\overline{\Braket{\hat{\sigma}_x}}$ for an XXZ Heisenberg model with $\Delta = -0.7$ as a function of rescaled time $t \, / \, \tau$. A stretched exponential fit is applied and the residuals are shown in \textbf{(b)}. \textbf{(c)} $\log\overline{\Braket{\hat{\sigma}_x}}$ as a function of  $t \, / \, \tau$ is shown for different $\Delta$ where both axes are in logarithmic scale. A pure exponential decay and the limit of the Ising model are added. For numerical details see Appendix \ref{Supp:disorder}.}
\label{figure1}
\end{figure}

Considering the particular case of the XXZ Hamiltonian  we perform numerical simulations using the dTWA method to obtain the dynamics of the global magnetization. Exemplarily, the observed decay of the magnetization for $\Delta = -0.7$ is shown in Fig.~\ref{figure1}(a) (blue curve). Additionally, a stretched exponential law was fitted yielding nearly perfect agreement (red dashed line). Note that we have rescaled time with the fitted decay time $\tau$. Fig.~\ref{figure1}(b) shows the residual of the applied fit to the data which lies in the percentage regime. 
We explore different regimes of anisotropy in  Fig.~\ref{figure1}(c) showing the dynamical behavior of $\log\overline{\Braket{\hat{\sigma}_x}}$ for $\Delta = \{-2.0, \, -0.7 , \, 0.0 , \, 0.7 , \,2.0\}$ (For relaxation dynamics including the fit similar to Fig.~\ref{figure1}(a), see Appendix \ref{additionalfits}). In this double logarithmic plot a stretched exponential law corresponds to a line with slope $- \beta$. The comparison to an exponential decay ($\beta = 1$, black dotted line) shows that the decay is clearly subexponential in all cases.
 The linear behavior over five decades indicates the suitability of a stretched exponential law to describe the relaxation dynamics. By comparing the dynamics of the various realizations of XXZ Heisenberg models to the analytical result of the Ising limit $\Delta \rightarrow \pm \infty$ (black dashed line), we observe remarkably good agreement.

Next, we investigate the stretched exponential relaxation of the system systematically over a broad range of values of the anisotropy parameter $\Delta$. For this we extract the two characteristic parameters $\beta$ and $\tau$ from the fit to the relaxations dynamics  of magnetization and purity for each value of the anisotropy parameter. The stretch power is largely independent of disorder strength in the regime of strong disorder and convergence with respect to system size is investigated as we show in detail in Appendix~\ref{Supp:disorder}.

\begin{figure}[h]
\includegraphics[width= \linewidth]{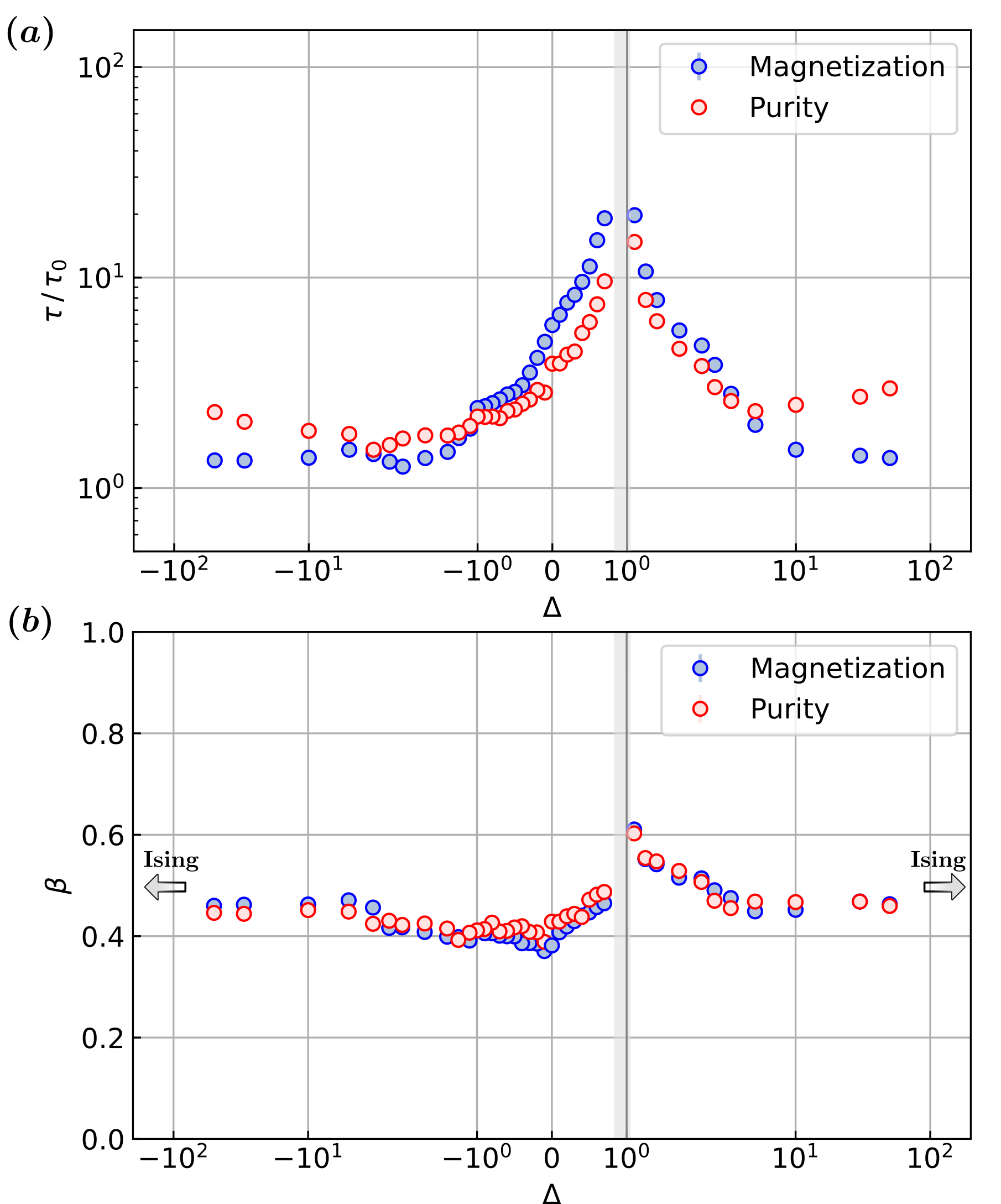}
\caption{The fit parameters $\tau \, / \, \tau_0$ \textbf{(a)} and $\beta$ \textbf{(b)} are shown for magnetization and ensemble-averaged single-spin purity as a function of anisotropy $\Delta$. For numerical details see Appendix \ref{Supp:disorder}. Within the grey shaded area, relaxation dynamics are not properly described by stretched exponential law.}
\label{figure2}
\end{figure}

The relaxation time $\tau$ of the magnetization (blue dots) and purity (red dots) is shown in Fig.~\ref{figure2}(a) as a function of $\Delta$. The y-axis is rescaled by the timescale $\tau_0$ corresponding to the magnetization's decay in the thermodynamic limit of the Ising model \cite{schultzen2021glassy}. The figure shows that this limit is well recovered for large $|\Delta|$ as the extracted $\tau$ lies close to $\tau_0$. Similar behavior is observed for the purity, where the value of $2 \tau_0$ is expected in the Ising limit. Remaining discrepancies are attributed to finite system size.
By approaching $\Delta = 1$, the observed dynamics for both quantities  is remarkably slowed down compared to the Ising case. This behavior is expected since no dynamics occur in the fully isotropic Heisenberg model. We point out the fact, that within the region closer to $\Delta = 1$, represented by the gray shaded area, relaxation dynamics become prohibitively slow. As a result, magnetization and purity have not fully relaxed at numerically accessible
times t, and the dynamics is not captured well by stretched exponential law. This region is therefore excluded from further discussion.

In Fig.~\ref{figure2}(b) the corresponding stretch powers for both purity and magnetization are shown. Remarkably, although timescales vary strongly as a function of $\Delta$, resulting stretch powers are within a narrow range (from $\beta  =0.4$ to $\beta = 0.6$) around the Ising value of $\beta = 0.5$. This indicates, that the glassy property of the relaxation, characterized by the stretch power $\beta$, is not strongly dependent on the underlying anisotropy and is similar for systems possessing different conservation laws. 

\section{Generalization to XYZ Heisenberg models}\label{section3}
We further explore the generality of the stretched exponential law by investigating a more general class of spin models, where the U(1) symmetry corresponding to conservation of the z-magnetization is broken. Specifically, we relax the constraint of $J_x=J_y$, thus realizing an XYZ model.
As a particular example, we set $J_x= 0.5$ and $J_y=1$ and vary $\Delta$. Another example, realizing a YZ Heisenberg model, where $J_x= 0$ is presented in Appendix \ref{YZcase}. 

\begin{figure}[h]
\includegraphics[width= \linewidth]{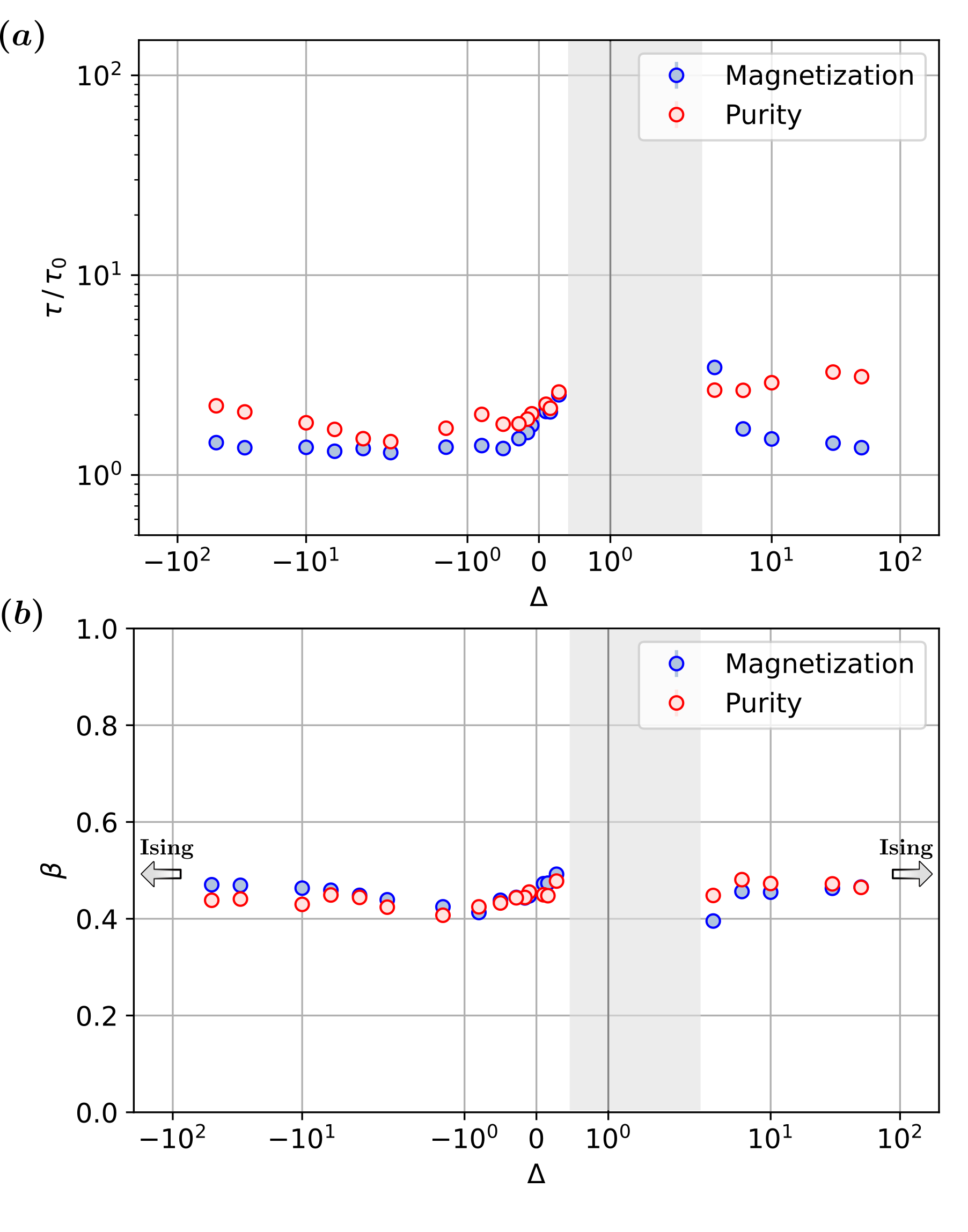}

\caption{The fit parameters $\tau \, / \, \tau_0$ \textbf{(a)} and $\beta$ \textbf{(b)} are shown for magnetization and ensemble-averaged single-spin purity as a function of $\Delta$ for an XYZ Hamiltonian with $J_x=0.5$, $J_y = 1$. For numerical details see Appendix \ref{Supp:disorder}. Within the grey shaded area, relaxation dynamics are not properly described by stretched exponential law.}
\label{figure3}
\end{figure}

The characteristic parameters of the fit are displayed in Fig.~\ref{figure3} in analogy to Fig.~\ref{figure2}.
The resulting behavior of $\beta$ and $\tau$ are similar to the results of Fig.~\ref{figure2}. One main difference is that the region around $\Delta = 1$ where we can not extract reliable quantities is broader than in the XXZ case. Nevertheless, the obtained stretch powers are still close to $\beta = 0.5$, confirming that breaking the symmetry does not alter drastically glassy relaxation behavior of the system.

\section{Conclusion}\label{section4}
Motivated by experimental findings of glassy dynamics in the XXZ Heisenberg model for $\Delta \approx -0.7$ and the analytical observation of stretched exponential law in the quantum Ising model, we have systematically investigated in this work the magnetization and ensemble-averaged purity dynamics of XYZ Heisenberg models. In all investigated cases, glassy dynamics characterized by stretched exponential relaxation are observed. We found that independent of the symmetries and Hamiltonian parameters, the stretch power lies close to the one analytically predicted for the Ising limit ($\beta = 0.5$). 

The numerical investigations in this work have been carried out using the semiclassical dTWA method, allowing us to simulate hundreds of particles in the strong disorder region. The application of dTWA is justified, since it is expected to succesfully approximate dynamics of one-point correlations \cite{PhysRevX.5.011022}. In addition, the method succeeded in the past to reproduce experimentally observed relaxation dynamics in disordered Heisenberg models \cite{PhysRevLett.120.063601,Signoles2021}. 
The obtained results on disordered XYZ models can be experimentally tested, as recent developments on Floquet engineering of Rydberg spins have enabled the possibility to experimentally implement Heisenberg Hamiltonians with tunable interaction coefficients \cite{geier2021floquet}.

Our observations of stretched exponential law in both XXZ and XYZ Heisenberg systems possessing a scale-invariant distribution of interaction strengths extend the connection between scale-invariance and glassy dynamics beyond the quantum Ising model. Furthermore, the presented work shows that this connection also constitutes a reality for non-integrable quantum spin system.

\section*{Acknowledgments}
This work is supported by the Deutsche Forschungsgemeinschaft (DFG, German Research Foundation) under Germany’s Excellence Strategy EXC2181/1-390900948 (the Heidelberg STRUCTURES Excellence Cluster), within the Collaborative Research Center SFB1225 (ISOQUANT) and the DFG Priority Program 1929 “GiRyd” (DFG WE2661/12-1). 
We acknowledge support by the European Commission FET flagship project PASQuanS (Grant No. 817482) and by the Heidelberg Center for Quantum Dynamics. C.H. acknowledges funding from the Alexander von Humboldt foundation and T.F. from a graduate scholarship of the Heidelberg University (LGFG). The authors acknowledge support by the state of Baden-Württemberg through bwHPC and the German Research Foundation (DFG) through grant no INST 40/575-1 FUGG (JUSTUS 2 cluster).

\bibliography{references, refs2}
\appendix
\section{Discrete Truncated Wigner Approximation}\label{dtwa}
\begin{figure}[t]
\includegraphics[width= \linewidth]{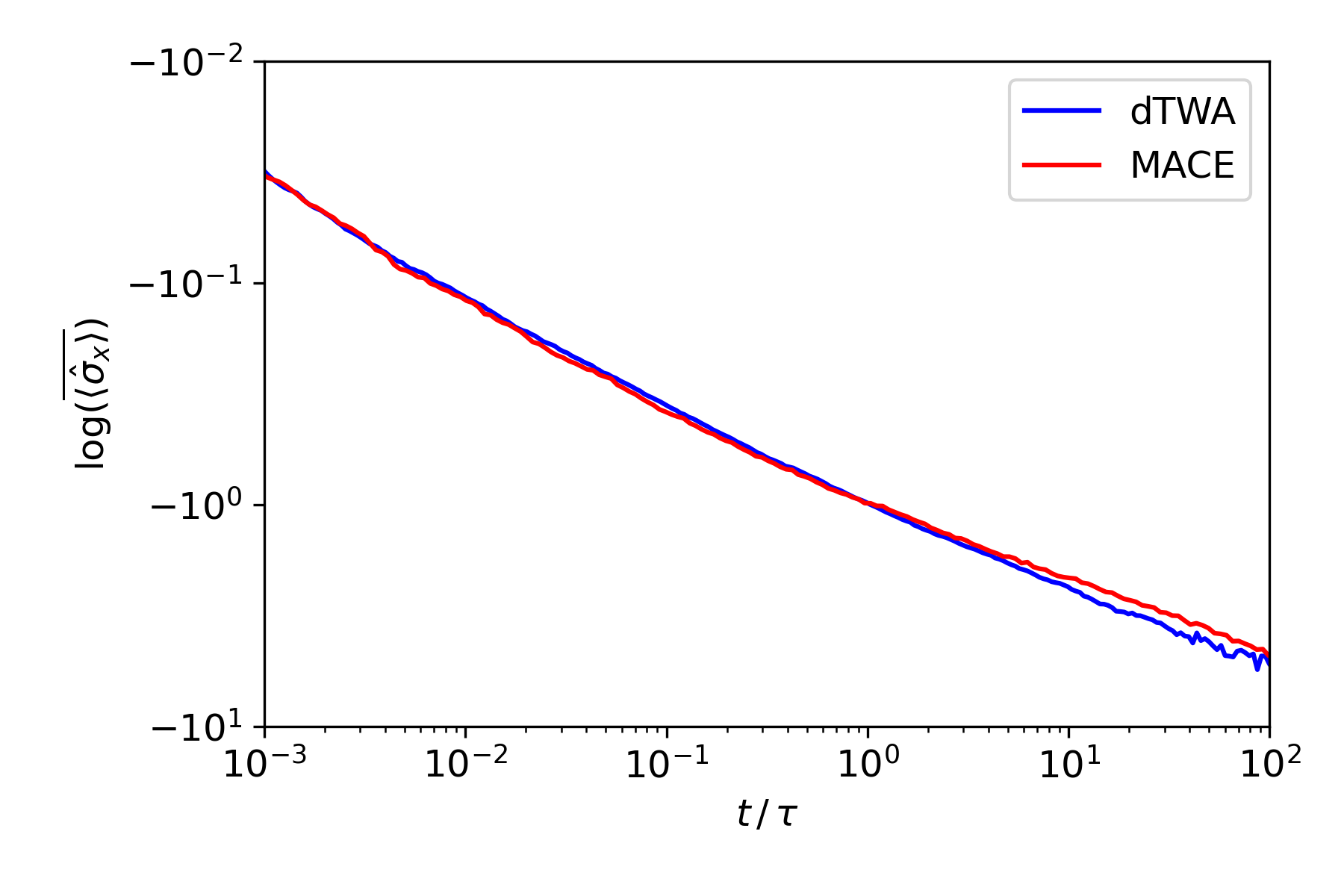}
\caption{$\log\overline{\Braket{\hat{\sigma}_x}}$ as a function of  $t \, / \, \tau$ for both dTWA and MACE for a XXZ Heisenberg model with anisotropy $\Delta = -0.7$. For the MACE simulation, clusters of 12 surrounding spins are chosen and the same spin positions were used. For these numerical simulations, we chose $N=100$ and $x= 8 \cdot 10^{-3}$ (See Appendix \ref{Supp:disorder}).}
\label{figureMace}
\end{figure}

dTWA can be understood as a Monte-Carlo average over trajectories sampled from the Wigner distribution of the systems initial state and evolved according to the mean-field equations of motions, thus accounting for quantum fluctuations \cite{PhysRevX.5.011022}. dTWA has been used succesfully to model the dynamics of one- and two-point correlations for the Ising model and the XY Heisenberg model \cite{PhysRevX.5.011022} and with experimental observations of the decay of transversal magnetization in Heisenberg models \cite{PhysRevLett.120.063601,Signoles2021}. In addition, it has been shown, that dTWA is capable of describing genuine quantum features like entanglement \cite{Lepoutre2019} and correctly reproduces the diffusive long time dynamics of generic non-integrable quantum systems\cite{Schuckert_2020}. 
While these arguments support to the applicability of dTWA in the system under study, we compare in the following the resulting dynamics to another numerical method based on exact diagonalization. The Moving Average Cluster Expansion (MACE) \cite{Hazzard_2014} solves clusters of spins exactly. This method takes the full quantum dynamics into account, as long as the correlation length is smaller than the cluster size. 
We present in Fig.~\ref{figureMace} relaxation dynamics of the transversal magnetization $\overline{\Braket{\hat{\sigma}_x}}$ for both methods over five decades for an XXZ Heisenberg model with $\Delta = -0.7$. Remarkably good agreement is found, thus cross validating both methods. In the work of this paper, we decided to use dTWA as it requires less computational resources, which allows us to explore a broader parameter space. In addition to the averaging performed via the Monte-Carlo trajectories, we perform disorder averages over different spin configuration drawn from the same uniform distribution. We point out, that the dynamics of the magnetization are  converged for 20 trajectories and 200 disorder averages, whereas for the purity it requires 200 trajectories and 40 disorder averages. 

\section{Dependence on system size and disorder strength}\label{Supp:disorder}
\begin{figure}[t]
\includegraphics[width= \linewidth]{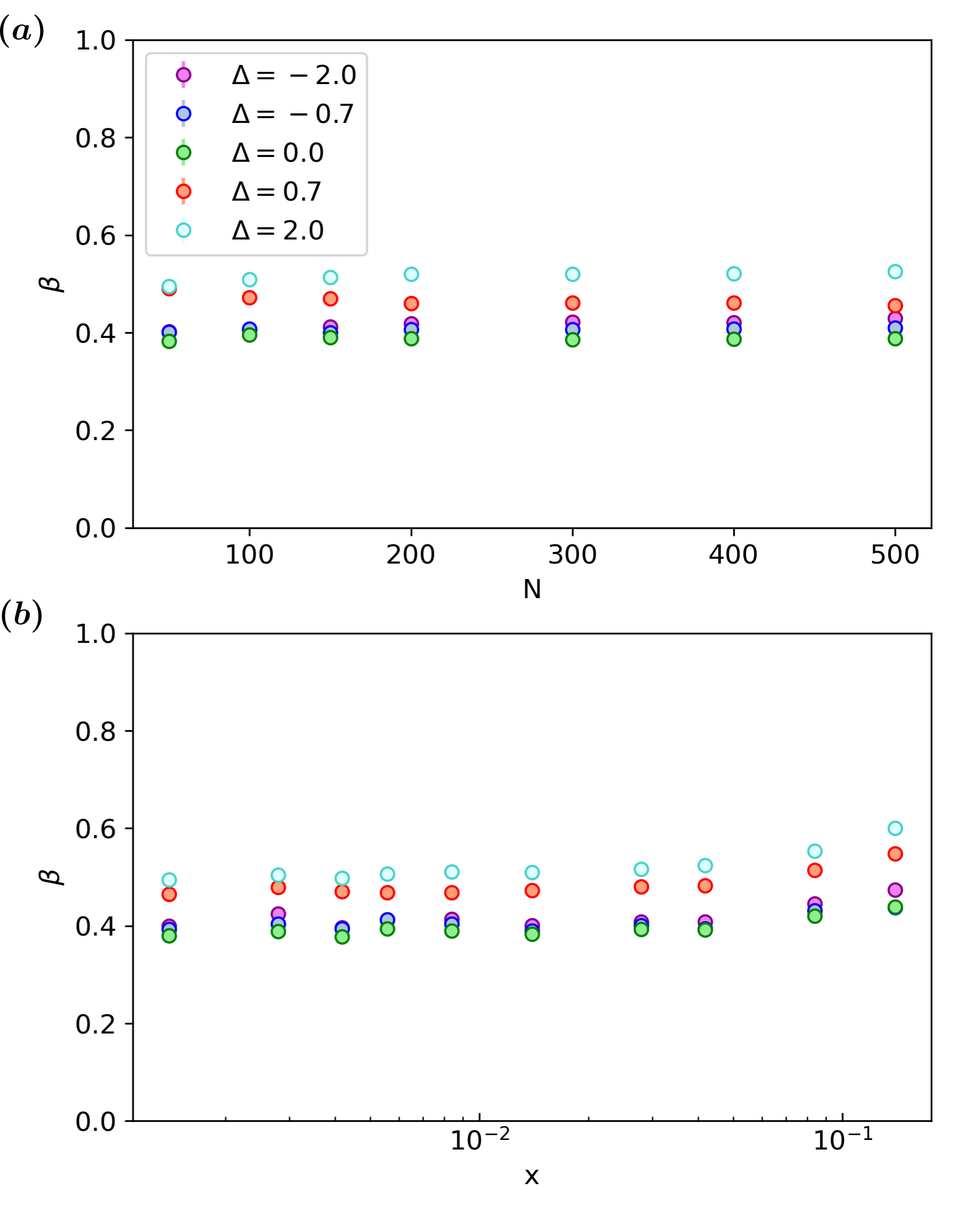}
\caption{$\beta$ exponents obtained from a stretched exponential fit to the numerically simulated transverse magnetization dynamics for different anisotropies  $\Delta$ as a function of \textbf{(a)} particle number $N$  and \textbf{(b)} disorder strength $x$ .}
\label{figureDisorder}
\end{figure}

In the whole paper we consider a system of $N$ spins drawn from a homogeneous spatial distribution in three dimensions characterized by the density $n$. Motivated by the realization of spin systems with Rydberg atoms where the blockade effect imposes a minimal distance between Rydberg spins, we consider a minimal distance $r_b$ between different spins. Disorder is coming from the random spin positions resulting in a broad distribution of interaction strengths. As a disorder quantifier we define $x = n V_b$, where $V_b$ is the spherical volume with radius $r_b$ \cite{schultzen2021glassy}. 
For large $x$, the induced correlations of $r_b$ play a significant role, and a more regular structure is obtained due to only slightly varying distances between the spins. For sufficiently small $x$, the induced correlations can be neglected and uncorrelated spin positions are reobtained, which justifies the choice of $x$ as the disorder parameter.\\ Additionally, we perform a systematic investigation of the scaling of the numerical results with particle number $N$ in order to assure, that remaining finite size effects are negligible in comparison to the structure obtained in Fig.~\ref{figure2}(b). In the following we focus on the glassy dynamics of the transverse magnetization. In Fig.~\ref{figureDisorder}(a), we see that the stretched power $\beta$, for a given $\Delta$, is only slightly dependent on the particle number between 50 and 500. No global trends are visible and remaining fluctuations resulting from finite size effects are varying from 0.01 ($\Delta = 0.0$) to 0.03 ($\Delta = 0.7$). Note, that for the Ising limit, the resulting stretch power for $N=100$ particles varies by 0.03 from the analytical result of the thermodynamic limit \cite{schultzen2021glassy}, which can also be seen in the large $|\Delta|$ limit of Fig.~\ref{figure2}(b) and Fig.~\ref{figure3}(b). Nevertheless, remaining finite size artefacts are negligible in comparison to the obtained values and structure of these figures. Therefore, we fix $N=100$ for all simulations shown.\\
Glassy relaxation has been found to be a robust feature for disorder strengths above a certain threshold \cite{Signoles2021}. To ensure that the chosen disorder parameter $x$ is within his universal strong disorder regime we investigate how the stretch power $\beta$ changes when we actively modify the disorder strength by varying $x$. In Fig.~\ref{figureDisorder}(b), we show the resulting stretch powers for different $\Delta$ values in the case of an XXZ Hamiltonian. We see that if $x$ lies below a value of 0.05, the $\beta$ exponent obtain from the fit to the dynamics gets disorder independent. Within this paper, We have chosen to work in this limit by choosing $x= 8 \cdot  10^{-3}$.

\section{Additional agreement of stretched exponential law}\label{additionalfits}
\begin{figure}[h]
\includegraphics[width= \linewidth]{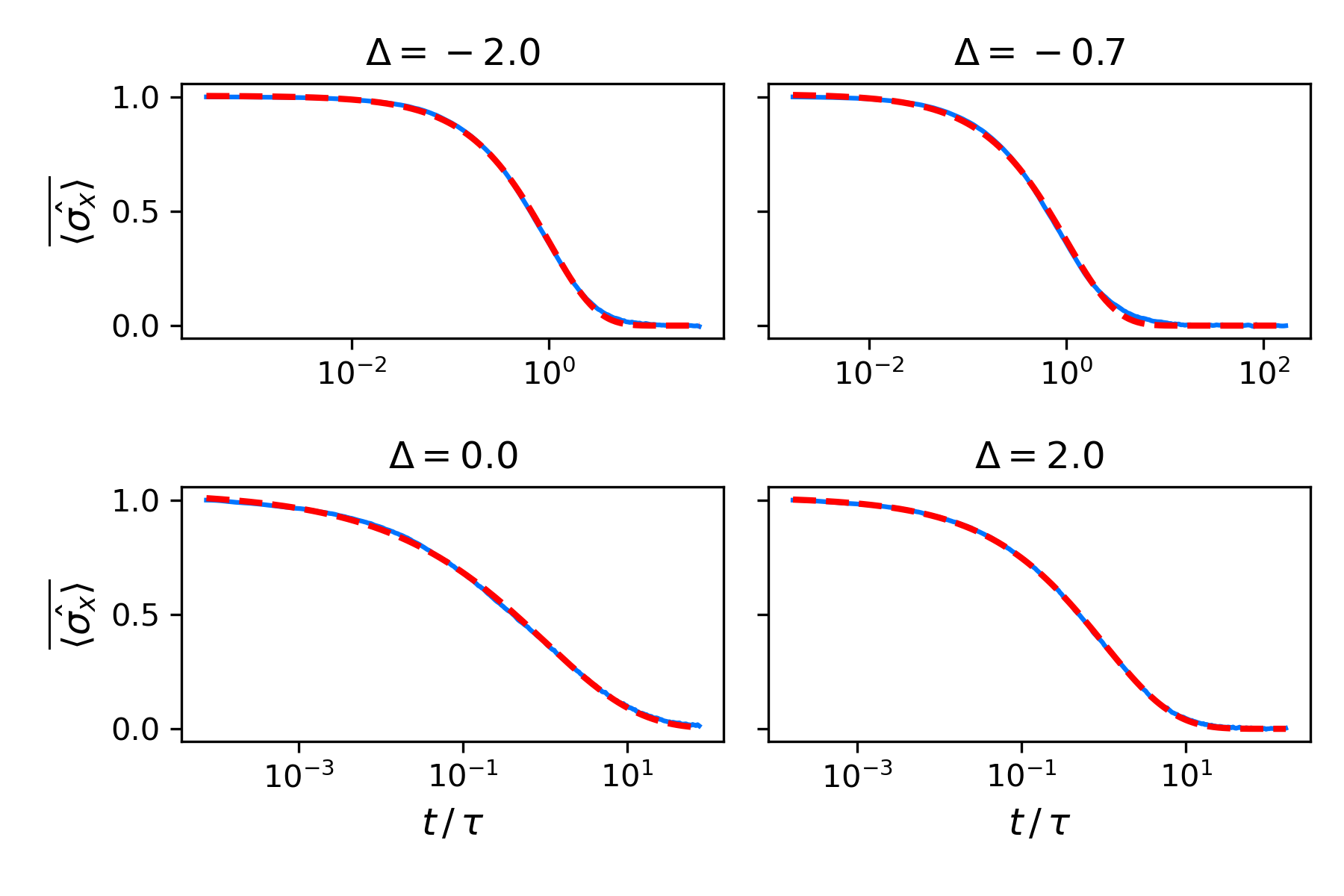}

\caption{Transverse magnetization as a function of $t \, / \, \tau$ for different anisotropies. Numerical data is obtained for $N=100$ and $x= 8 \cdot  10^{-3}$ (See Appendix \ref{Supp:disorder}).}
\label{figureotherfits}
\end{figure}

Complementary to the data presented in Fig.~\ref{figure1}(a) we provide the relaxation dynamics of the transverse magnetization with linear y-scale for the remaining anisotropies of Fig.~\ref{figure1}(c) $\Delta = \{ -2.0, 0.0, 0.7, 2.0\}$ in Fig.~\ref{figureotherfits}. We highlight the remarkable agreement between numerical data and the stretched exponential law. 

\section{Investigation on a YZ Heisenberg Hamiltonian}\label{YZcase}

\begin{figure}[h]
\includegraphics[width= \linewidth]{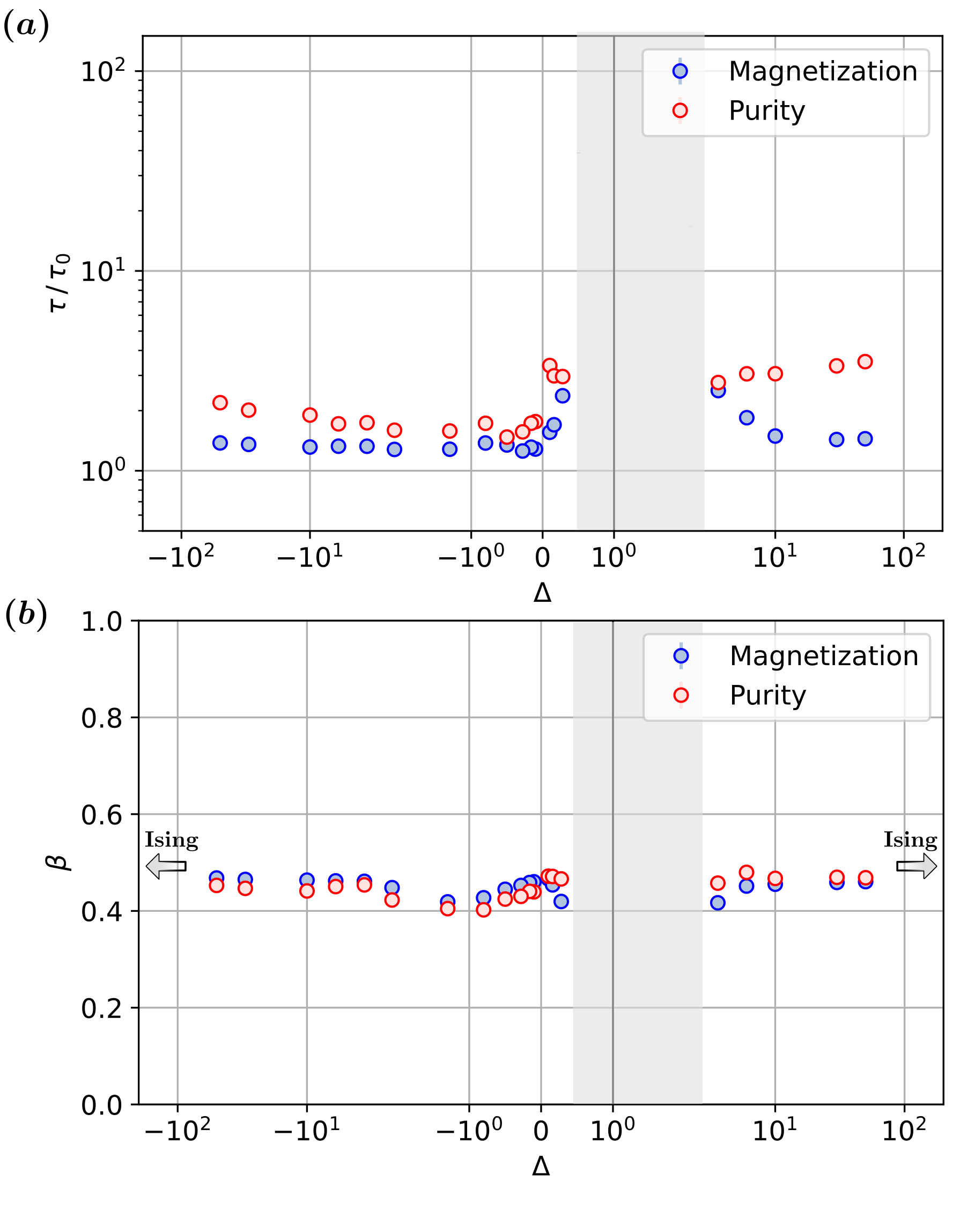}

\caption{The resulting fit parameters $\tau \, / \, \tau_0$ \textbf{(a)} and $\beta$ \textbf{(b)} are shown for magnetization and ensemble-averaged single-spin purity as a function of $\Delta$ for an XYZ Hamiltonian with $J_x=0$, $J_y=1$. For numerical details see Appendix \ref{Supp:disorder}. Within the grey shaded area, relaxation dynamics are not properly described by stretched exponential law.}
\label{figureYZ}
\end{figure}

We present in Fig.~\ref{figureYZ} a complementary plot to the one presented in Fig.~\ref{figure3} where this time $J_x=0$ and $J_y = 1$. The resulting fit parameter are similar to the ones presented in Fig.~\ref{figure3}.
\end{document}